%% file: cerium.tex
\UseRawInputEncoding
\documentclass[aps,preprint,llncs]{revtex4}
\usepackage[utf8]{inputenc}
\usepackage[english]{babel}
\usepackage[T1]{fontenc}
\usepackage{amsmath}
\usepackage{amsfonts}
\usepackage{ragged2e}
\usepackage{amssymb}
\usepackage{xcolor}
\usepackage[colorlinks=true,linkcolor=blue,filecolor=blue, urlcolor=blue,citecolor=blue]{hyperref}
\usepackage{graphicx}
\usepackage{float}
\usepackage{color}
\usepackage{soul}
\usepackage{siunitx}
\usepackage{physics}
\usepackage{lineno}
\usepackage{subfiles}

\begin{document}

\subfile{myfirstfile}

\pagebreak
\subfile{mysecondfile}

\end{document}

%% file: myfirstfile.tex
\title{Above 20K conventional superconductivity in Cerium}

\author{Mohd Monish, Nikhlesh S. Mehta, Mona Garg}
\author{Goutam Sheet}

\email{goutam@iisermohali.ac.in}

\affiliation{Department of Physical Sciences, Indian Institute of Science Education and Research (IISER) Mohali, Sector 81, S. A. S. Nagar, Manauli, PO 140306, India}

\begin{abstract}


A high superconducting critical temperature ($T_c$) under normal laboratory conditions in a material that is chemically simple and stable, like an elemental metal, is a hitherto unattained goal of modern science and technology. Certain elemental metals are known to display reasonably high $T_c$ only under extraordinarily high pressures where their spectroscopic characterization and application are tightly restricted. Here we show that a $T_c$ exceeding 20 K can be realized on pure elemental Ce under uniaxial pressure created simply by pressing a sharp metallic needle on the metal. This is a breakthrough because pure Ce doesn't superconduct under ambient conditions and the application of 54 GPa of hydrostatic pressure yields only a low $T_c$ of 1.8 K in the metal. In addition, by driving the area under the needle in a mechanically controlled way to the ballistic transport regime, for the first time, we spectroscopically characterized the superconducting energy gap in a high-pressure superconducting phase and found that superconducting Ce respects the conventional Bardeen-Cooper-Shrieffer (BCS) theory.

\end{abstract}

\maketitle

\newpage
Superconductors hold the promise to reduce the energy needs of the future and develop novel technologies ranging from quantum computers, quantum sensors, medical imaging equipment and faster public transport systems.  It is known that in order to exploit the remarkable physical properties of a superconductor in real life, it is most important to stabilize a high-temperature superconducting phase in a simple material that is easily synthesized, low-cost and remains stable in air. In the past few decades, application of high pressure has been used as an effective tool for realizing novel superconducting phases with very high critical temperatures ($T_c$) and for tuning their superconducting properties \cite{Zhang2017}. Such phases usually emerge due to a structural modification under high pressure which facilitates non-superconductors or even insulating solids to exhibit superconductivity \cite{densedlithium,Struzhkin1997,Shimizu2002,Shimizu1998,PhysRevLett.121.037004, 
PhysRevB.61.R3823,PhysRevB.78.064519,Ti2,HAMLIN200782,SHIMIZU201830}. Recent discoveries have shown that certain complex materials, like the superhydrides, exhibit very high-temperature superconductivity at extremely high hydrostatic pressures of several hundreds of Giga Pascal (GPa) \cite{superhydride,LaH10,CeH10,BI2019}. Certain elemental metals, e.g., calcium, titanium and scandium are known to display unexpectedly high critical temperatures under an increased pressure inside specially designed pressure cells \cite{calcium,Titanium,Scandium,scandium1,scandium2}. However, in all these cases, the requirement of unusually high pressure and the specially designed pressure cells poses a formidable challenge which limits the fundamental spectroscopic characterization and industrial application of such novel superconductors. In this work we present a simple, easily affordable technique to acquire superconductivity with $T_c$ exceeding 20 K in a pure elemental metal Cerium (Ce) under easily attainable conditions in a laboratory where we also successfully carried out a spectroscopic investigation of the high-pressure phase.

\begin{figure}[h!]
\centering
\includegraphics[scale = 0.58]{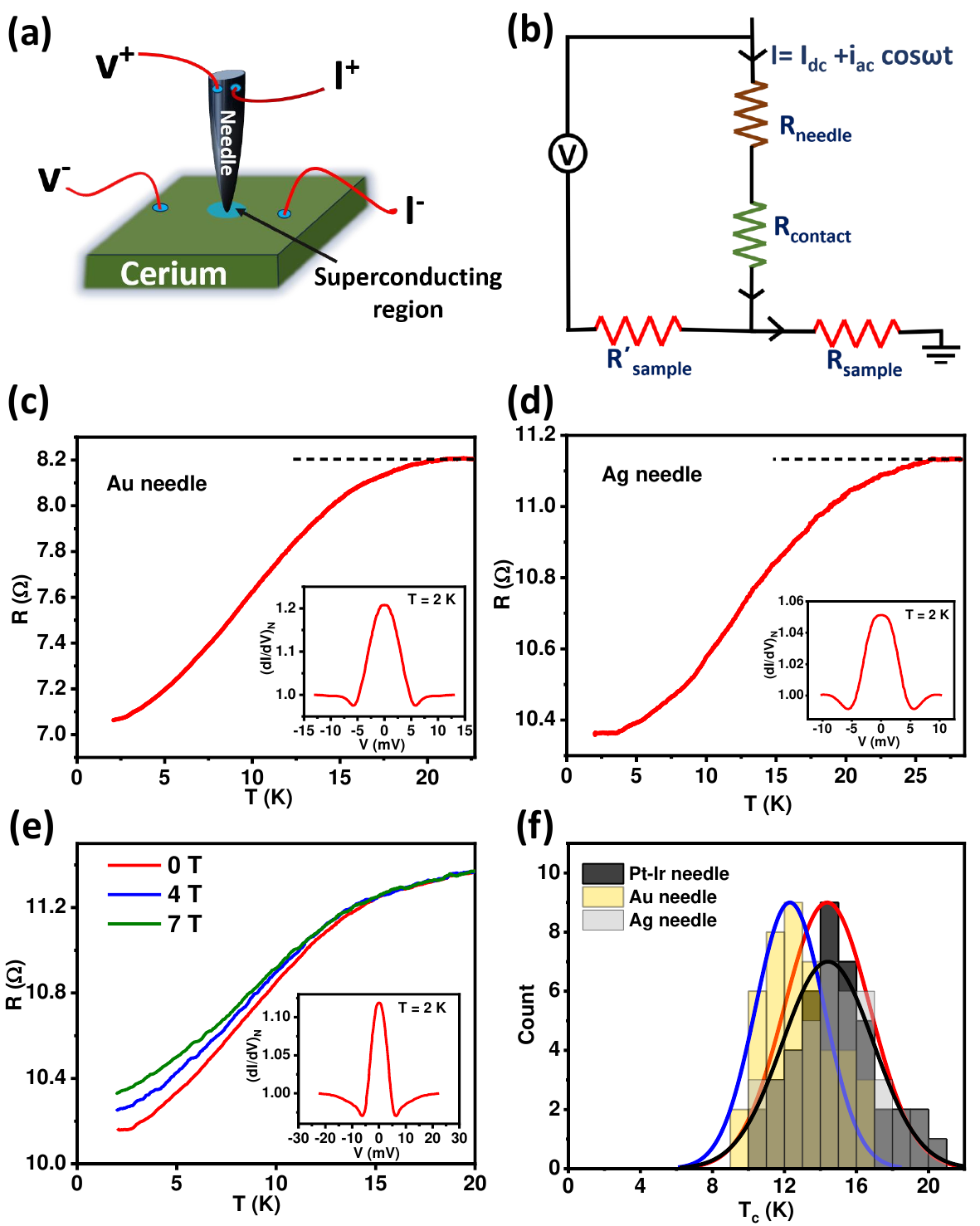}
\caption{\textbf{(a) A schematic diagram of the measurement setup. (b) The measured resistances. $R_{contact}$ is larger than the others. (c) Resistance ($R$) vs. temperature ($T$) of a contact between Ce and Au needle. The $inset$ displays the corresponding conductance spectrum. (d)$R$ vs. $T$ of a contact between Ce and an Ag needle. The $inset$ shows the corresponding conductance spectrum. (e) $R$ vs. $T$ of the point contact at different magnetic fields ($H$) up to 7 T. The inset shows the corresponding conductance spectrum, (f) Distribution of $T_c$ for 130 different contacts.}}
\label{fig1}
\end{figure}

Cerium (Ce) is an element known for displaying a range of unusual properties and is special for being the first element in the periodic table having 4$f$ electrons \cite{PhysRevB.99.045122,Wu2021}. The electronic configuration of Ce is [Xe] 4$f^1$5$d^1$6$s^2$. Since the energies of the 4$f$, 5$d$ and 6$s$ are very close, small changes in energy causes drammatic changes in the electronic structure of Ce. This variability is thought to be responsible for the outstanding electronic properties as well as the existence of a variety of allotropes ($\alpha$, $\beta$, $\gamma$, $\delta$ etc.) of Ce \cite{Nikolaev_2012}. 
At ambient pressure and room temperature, Ce crystallizes in a face-centered cubic $\gamma$ phase and is known for structural transitions induced by pressure and temperature \cite{Munro_2020}. Upon applying pressure, Ce undergoes a series of structural transitions: $\gamma$ phase to isostructural $\alpha$ phase \cite{PhysRev.76.301,Johansson4,PhysRevLett.49.1106}, followed by either the $\alpha^\prime$ phase (orthorhombic) or $\alpha^{\prime\prime}$ phase (monoclinic), depending on the sample preparation procedure \cite{PhysRevLett.78.3884,PhysRevLett.32.773}, and finally transforming into the body-centered tetragonal $\epsilon$ phase \cite{Vohra}. At room temperature, the transition from $\gamma$ to $\alpha$ occurs at around 0.8 GPa and the $\epsilon$ phase appears above 15 GPa of hydrostatic pressure. The dominant stable phase at liquid helium temperatures is the volume-collapsed $\alpha$ at ambient pressure. At such temperatures, under high hydrostatic pressure, the system is thought to go into further structural phases that remain to be fully characterized. As per ref \cite{Zhang} superconductivity emerges above 5 GPa in Ce with a maximum $T_c$ of 1.25 K which is achieved at a pressure of 17.2 GPa. Beyond this pressure, $T_c$ is seen to monotonically decrease and becomes $\sim$ 500 mK at 55 GPa. In another study, a $T_c$ of 1.8 K in Ce was achieved at 5 GPa which monotonically decreased with further increase in pressure \cite{Probst1975}. These results indicate that the upper bound of the pressure-induced $T_c$ in Cerium is 1.8 K. In this work, we have broken this limit by applying uniaxial pressure and achieved an order of magnitude higher $T_c$ in Ce.


We stabilized a superconducting phase of cerium with a remarkably high $T_c >$ 20 K using a simple needle-anvil technique involving no expensive, special instrumentation where a sharp metallic needle is pressed on pure Ce using a piezo motor. We tested this with the needles of three metals: Ag, Au, and Pt-Ir and the superconducting phase was obtained for all. We surmise that due to the small tip-area of the needle, even a small amount of force can generate a large pressure locally, giving rise to a (uniaxial) pressure-induced high-temperature superconducting phase in Ce. This phase was not detected earlier under hydrostatic pressure in Ce. This suggests that the uniaxial pressure provides access to the structural and electronic phases in Ce that are distinct from the phases obtained under hydrostatic pressure. In such phases the role of the 4$f$ electron in Ce might be unprecedented and could give rise to the observed high $T_c$ superconductivity under uniaxial pressure. Furthermore, the experimental setup allowed us to directly measure the superconducting energy gap by $\textit{in-situ}$ Andreev reflection spectroscopy \cite{andreev1964, Daghero_2010}. Thus, this report opens up a new area of research where a regular metal needle can be used to explore the emergence of uniaxial-pressure induced high-temperature superconductivity and their nature in simple, stable materials, as well as in complex quantum systems, without expensive pressure cells and allied instrumentation.

In Figure~\ref{fig1}(a), we have shown a schematic diagram describing how the superconducting phase was obtained by pressing a metallic needle on elemental cerium (Ce). In order to measure the electrical transport properties of the area under the needle, we employed a four probe technique by fixing the current and the voltage electrodes as shown in the schematic. In this geometry, as shown in Figure~\ref{fig1}(b) we measure a contact resistance between Ce and the needle, along with the resistance of the needle and a part of the sample that takes part in effective electric contact formation. Among these, the contact resistance is dominating and it is of the order of $\sim$ 10 $\Omega$ while the others are in the m$\Omega$ range. Two representative resistance ($R$) vs. temperature ($T$) graphs of the contact between an Au and Ag needle and Ce are shown in Figure 1(c) and Figure 1(d) respectively. The downward fall of $R$ is clearly seen below 20 K. As expected for a superconducting transition, the transition temperature decreases with increasing magnetic field (Figure 1(e)) \cite{tinkham1975introduction}. We have recorded data by pressing the needles at 130 different points on Ce with three different metallic needles (Pt-Ir, Ag, and Au) and found a distribution of $T_c$ as shown in Figure~\ref{fig1}(f). While a gold needle yields a lower average $T_c$ of 12 K, both Pt-Ir and Ag needles give rise to a maximum $T_c$ exceeding 20 K. This is an order of magnitude higher than the $T_c = $ 1.8 K reported in cerium under hydrostatic pressure up to 55 GPa.

\begin{figure}[h!]
\centering
\includegraphics[scale = 0.63]{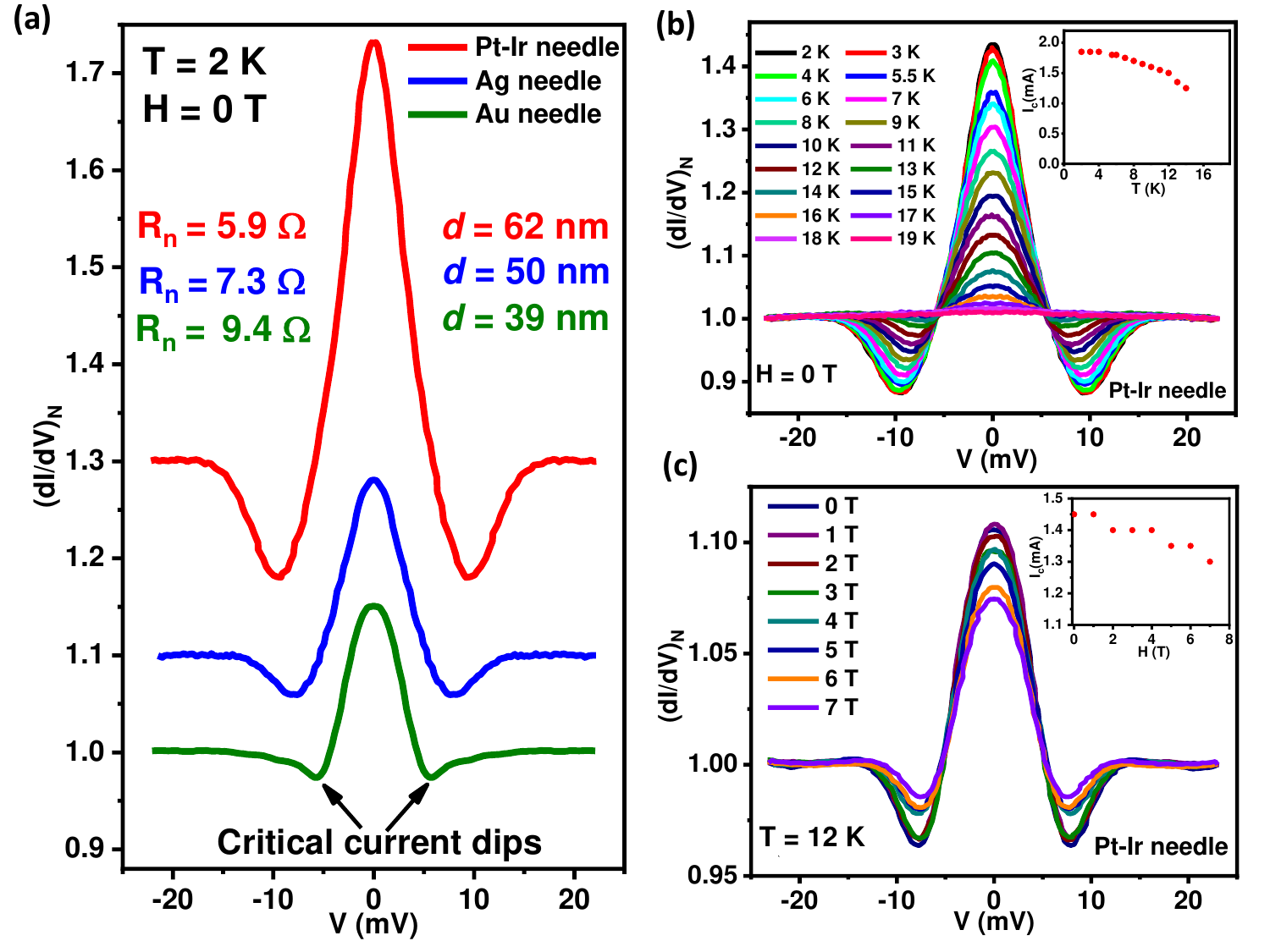}
\caption{\textbf{(a) Three representative normalized conductance spectra (with different needles and contact diameters (\textbf{$d$}) in the thermal regime showing critical current driven dips and low-bias conductance enhancement. (b) $T$ dependence of one spectrum in the thermal regime. The inset shows the $T$ dependence of critical current ($I_c$). (c) Magnetic field ($H$) dependence of the conductance spectrum recorded at 12 K. The inset shows the $H$ dependence of $I_c$.}}
\end{figure}

\begin{figure}[h!]
\centering
\includegraphics[scale = 0.6]{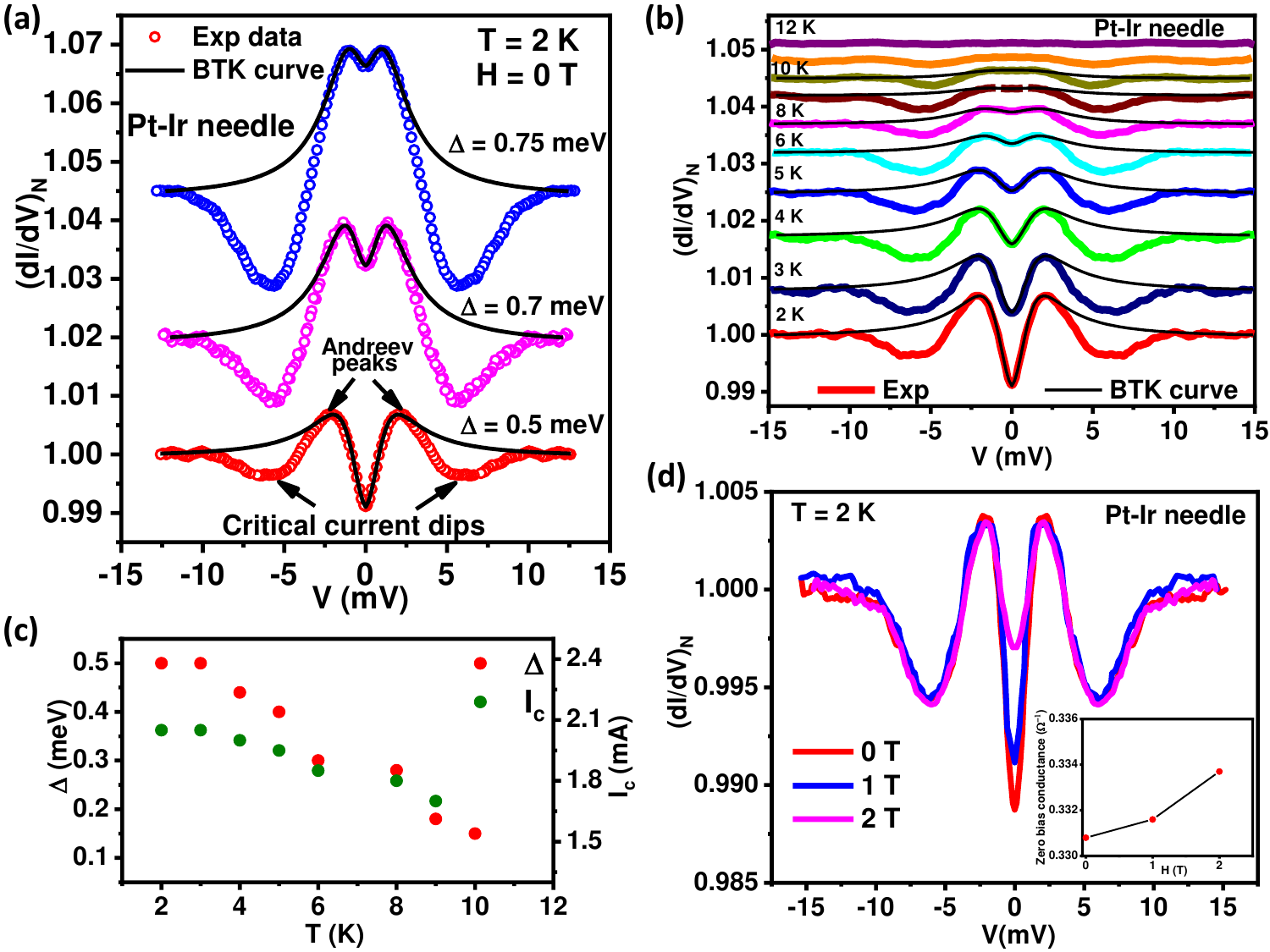}
\caption{\textbf{(a) Three representative normalized conductance spectra (colored circles) obtained in the intermediate regime showing dips due to critical current and peaks due to Andreev reflection. The solid black lines represent the best fits by BTK model. (b) $T$ dependence of a conductance spectrum obtained in the intermediate regime. (c) $T$ dependence of superconducting gap $\Delta$ and $I_c$. (d) Magnetic field ($H$) dependence of the conductance spectra. The inset shows the $H$ dependence of zero bias conductance.}}
\end{figure}

It is straightforward to note that the measurement geometry (Figure 1(b)) does not allow us to measure a zero resistance, but that does not limit us from unambiguously establishing the existence of superconductivity\cite{Aggarwal2016}. While the part of the sample that takes part in the contact formation goes to zero below $T_c$, the other resistors in the circuit remain non-zero. Nevertheless, for such mesoscopic superconducting phases, where a zero-resistance cannot be measured, there is an established recipe to confirm superconductivity by exploring the $dI/dV$ vs. $V$ characteristics in the different regimes of mesoscopic transport \cite{Aggarwal2016}. The insets of Figure 1(c,d,e) show the respective normalized $dI/dV$ vs. $V$ spectra at 2 K and zero magnetic field. The enhanced zero-bias conductance along with the sharp dips in $dI/dV$ closely resemble the spectra that are routinely recorded from point contacts between normal metals and superconductors, when the contact diameter is larger than the mean free path and the contact falls in the thermal regime of transport \cite{Goutam_criticalcurrent}. To illustrate the features with enhanced clarity, we presented three representative $dI/dV$ vs. $V$ thermal regime spectra acquired with three different needles and with different normal state resistance (Figure 2(a)). The lowest resistance spectrum (plotted in red) indicates the highest contact diameter and shows the maximum enhancement with sharper dips appearing at the highest $V$ (= $\pm$ 10 mV). The temperature dependence of the 5.9 $\Omega$ spectrum (Figure 2(b)) with Pt-Ir needle shows that the spectral features systematically evolve with temperature before they completely disappear at $T_c=$ 20 K. The magnetic field dependence at 12 K shows the suppression of the spectral features by a magnetic field. The position of the critical current driven dips move closer to zero with increasing temperature and magnetic field (insets of Figure 2(b,c)). All these collectively reconfirm that the spectral features are associated with an above 20 K superconducting phase of Ce realized under the needle.

Now it is imperative to explore the other regimes of mesoscopic superconductivity. The experimental setup allows us to modify the transport regime in a controlled way by operating the piezo-stage on which the needles are mounted. In Figure 3(a) we present spectra obtained between Ce and a needle of Pt-Ir in the intermediate regime (a transport regime that falls between the thermal and the ballistic regime). In this regime, we find the emergence of the double peak structures, symmetric about $V = 0$, along with the critical current driven dips. The double-peak structure is a hallmark signature of Andreev reflection between a normal metal and a superconductor. The black lines on top of the experimental data points show the simulations within the Blonder-Tinkham-Klapwijk (BTK) formalism \cite{BTK}. While the low-bias parts of the spectra are well described by the BTK model, the high-bias parts significantly deviate due to the thermal effects that involve the critical current\cite{Goutam_criticalcurrent,Aggarwal2016}. The features evolve systematically with temperature (Figure 3(b)) and magnetic field (Figure 3(d)) showing consistency with mesoscopic superconducting transport in the intermediate regime. In this regime, an estimate of the magnitude of the superconducting gap ($\Delta$) can be made from the BTK simulation of the low-bias parts of the spectra (Figure 3(c)). However, that is an underestimation due to the coexisting thermal effects. The minute shift in the position of the critical current dips in Figure 3(d) up to 2 Tesla indicates that the phase has a very high critical magnetic field. The Andreev reflection feature (zero-bias conductance dip), however, shows a pronounced magnetic field dependence.

\begin{figure}
\centering
\includegraphics[scale = 0.65]{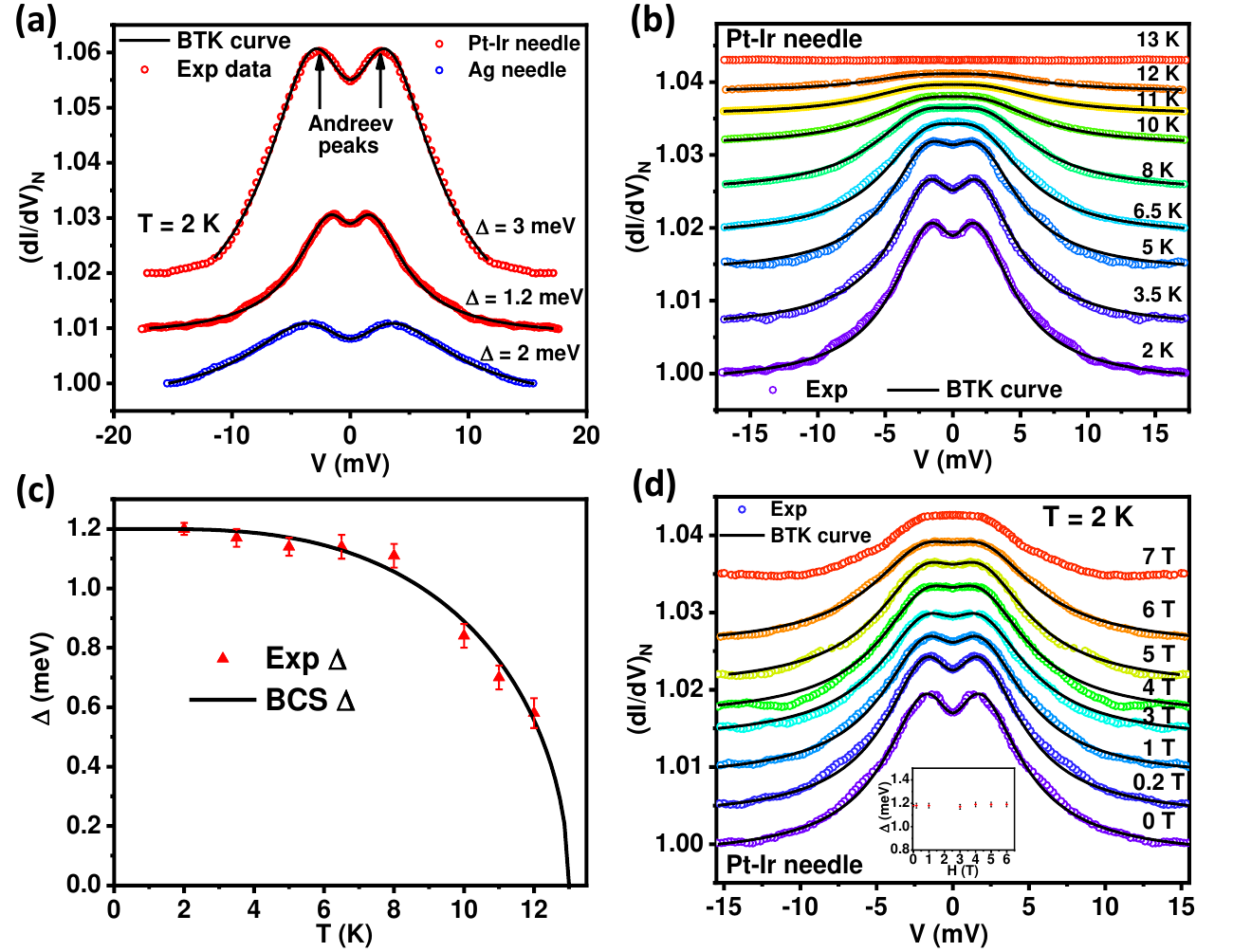}
\caption{\textbf{(a) Three representative normalized conductance spectra (colored circles) obtained in the ballistic regime showing the gap features. The solid black line represents the theoretical fits to each spectrum using BTK theory. (b) Temperature ($T$) dependence of the conductance spectra and corresponding BTK fits. (c) $T$ dependence of superconducting gap ($\Delta$) extracted from the spectra shown in Figure 4(b). (d) Magnetic field ($H$) dependence of the conductance spectra and corresponding BTK fits. The inset shows the variation of $\Delta$ with magnetic field ($H$).}}
\end{figure}

The experimental data presented above collectively confirm the emergence of an above 20 K superconducting phase on Ce under the metallic needles. To understand the mechanism, we note that the diameter of the contacts between the needles and Ce, as estimated using Wexler's formula \cite{Wexler_1966}, varied between 10 nm and 200 nm. We press the needle on the sample with an attocube walker which can take a maximum load of 5g. Even if we impart a force equivalent to 100 mg, ideally for a contact diameter of 100 nm, the pressure under the needle should be 100 GPa. At such high pressures, a local structural change is anticipated which might lead to superconductivity. The actual pressure under the needle is expected to be far less (of the order of a few hundred MPa), which is limited by the tensile strength of the metal, beyond which the needle becomes blunt in order to withstand the applied force. This idea is consistent with the $T_c$ distribution shown in Figure 1(f). Gold has the lowest tensile strength and Pt-Ir has the highest and consequently, the average $T_c$ is lowest for Au and highest for Pt-Ir. We also note that our observation of Ce superconductivity above 20 K cannot be directly compared with the experiments under high hydrostatic pressure. In this context we highlight the fact that the pressure applied by the needle is uniaxial in nature and as a consequence, we get access to a structural phase of Ce that is distinct from the known structural phases of Cerium under hydrostatic pressure. Within the limits of our experimental technique, this new phase remains structurally undetermined and requires further crystallographic characterization in order to gain a full understanding on the fate of the 4$f$ electron and how it can couple with the lattice to give rise to the remarkably high $T_c$ superconductivity. For completeness, we have also performed similar experiments on Scandium (Sc) where a hydrostatic pressure induced high $T_c$ superconducting phase was earlier reported. However, Sc did not superconduct under the needle (see section IX in SI). Also, a similar experiment on Lanthanum (La) detected only the known 5K superconductivity of the metal as seen under ambient pressure (see section X in SI). These observations support the idea that the 4$f$ electron in Ce plays a significant role behind the emergence of the superconducting phase under a needle. Depending on the shape of the needle, a pressure distribution under the needle is expected (see section I in SI) which also causes a local distribution of $T_c$. This is the reason behind the broad transition seen in Figure 1(c,d,e). Such a distribution also gives rise to a large broadening parameter ($\Gamma$) in the BTK analysis of the $dI/dV$ spectra \cite{PhysRevLett.93.156802}. This effect is discussed in detail in the section IV of SI.

Encouraged by the observation of the Andreev reflection features in the intermediate regimes, we attempted to establish a contact in the ballistic regime of transport where spectroscopic information could be achieved \cite{naidyuk2005point}. In Figure 4(a) we show three representative ballistic (or diffusive) regime spectra with clear double peak structures due to Andreev reflection and no non-ballistic transport characteristics \cite{Goutam_criticalcurrent,Ritesh}. In this regime, the Andreev peaks appear near $V = \Delta/e$. We also show the BTK fits as black lines. The superconducting gap features display a strong magnetic field dependence and they are almost suppressed at 7 Tesla (Figure 4(d). The fits of the different spectra yield a distribution of the superconducting energy gap, the highest being 3 meV. This yields $\Delta/k_BT_c \sim $ 1.76 indicating that the superconducting phase falls within the weak-coupling Bardeen-Cooper-Schriefer (BCS) limit \cite{BCS}. The remarkably good fits with the BTK theory and the estimated gap ratio indicate that the superconducting phase is conventional in nature. To further confirm this, we performed temperature dependence of the ballistic regime spectra. 
A representative temperature dependence along with the BTK fits are shown in Figure 4(b). The extracted $\Delta$ follow a BCS behavior (black line). 

In conclusion, we have discovered a superconducting phase above 20 K on the elemental metal Ce by pressing it with a metallic needle. The phase emerges due to the uniaxial pressure applied by the needle. We presented temperature and magnetic field dependent differential conductance spectra in different regimes of mesoscopic transport involving the new superconducting phase. In the ballistic regime, we performed spectroscopy to extract the magnitude of the superconducting energy gap as a function of temperature and magnetic fields. The analysis of the spectra confirms that the superconducting phase under the needle is well described by the conventional BCS formalism. The results not only provide a new recipe to realize high-temperature superconductivity through a low-cost, affordable route, it also, for the first time ever, demonstrates how spectroscopic information can be reliably extracted from such a high-pressure superconducting phase. 
\section*{Acknowledgements}
The authors acknowledge fruitful discussions with Prof. Tanusri Saha-Dasgupta. M.M. thanks the Council of Scientific and Industrial Research (CSIR), Government of India, for financial support through a research fellowship (Award No. 09/0947(12989)/2021-EMR-I). N.S.M. thanks University Grants Commission (UGC) for senior research fellowship (SRF). M.G. thanks CSIR, Government of India, for financial support through a research fellowship (Award No. 09/947(0227)/2019-EMR-I). GS acknowledges financial assistance from Science and Engineering Research Board (SERB), Govt. of India (grant number: \textbf{CRG/2021/006395}).

\bibliography{cerium}

%% file: mysecondfile.tex
\title{Supplementary Information\\ 
\Large Above 20K conventional superconductivity in Cerium}
\author{Mohd Monish, Nikhlesh S. Mehta, Mona Garg and Goutam Sheet}
\email{goutam@iisermohali.ac.in}

\affiliation{Department of Physical Sciences, Indian Institute of Science Education and Research (IISER) Mohali, Sector 81, S. A. S. Nagar, Manauli, PO 140306, India}

\maketitle

\justifying
\section{Distribution of pressure}
Consider two bodies pressed together by forces being applied in a direction perpendicular to their surfaces. We have considered no friction, axially symmetrical, and no contact deformation. Let us see the solution for the pressure distribution of four different contact profiles: flat, cone, spherical and parabolic indenters \cite{Pressure_distribution}. These four shapes form the ideal base shapes for most contacts in technical applications.

\subsection{Flat cylindrical punch}
Pressure distribution in the contact area is
\begin{equation}
    P(r,d) = \frac{E^{*}}{\pi \sqrt{a^2 - r^2}}, r\leq a
\end{equation}
where d is the indentation depth, a is the radius of the contact and $E^{*}$ is the effective elasticity modulus and is given by
\begin{equation}
    \frac{1}{E^{*}}= \frac{1-\nu^{2}_{1}}{E_1} + \frac{1-\nu^{2}_{2}}{E_2}
\end{equation}
where $\nu_1$, $\nu_2$ are Poisson's ratio and $E_1$, $E_2$ are elasticity moduli of the two bodies in contact (in our case Ce sample and metallic needle). 

\subsection{Conical indenter}
In the case of the conical indenter, the pressure distribution in the contact area is
\begin{equation}
    P (r,a) = + \frac{1}{2}  E^{*} 
  tan\theta \ln{\left(\frac{a+\sqrt{a^2-r^2}}{r}\right)}
\end{equation}

\subsection{Spherical indenter with profile radius $R$ and contact diameter $a$}
The pressure distribution in the contact area in the case of a spherical indenter with profile radius $R$ and contact diameter $a$ is
\begin{equation}
    P (r,a) = \frac{E^{*}}{\pi} \int_{r}^{a} \left[ \frac{xR}{R^2 - x^2} + \frac{1}{2} \ln{\left( \frac{R+x}{R-x} \right)} 
      \right]\frac{1}{x^2 - r^2} \,dx  , r\leq a
\end{equation}
 where $E^{*}$ is the effective elasticity modulus and is given by
\begin{equation}
    \frac{1}{E^{*}}= \frac{1-\nu^{2}_{1}}{E_1} + \frac{1-\nu^{2}_{2}}{E_2}
\end{equation}
\subsection{Paraboloid indenter}
The pressure distribution in the contact area in the case of a paraboloid indenter of radius $R$ making contact diameter $a$ is
\begin{equation}
    P(r,a) = \frac{2E^{*}}{\pi R} \sqrt{a^2 - r^2}
\end{equation}
\begin{figure}[h!]
    \centering
\includegraphics[width=0.85\linewidth]{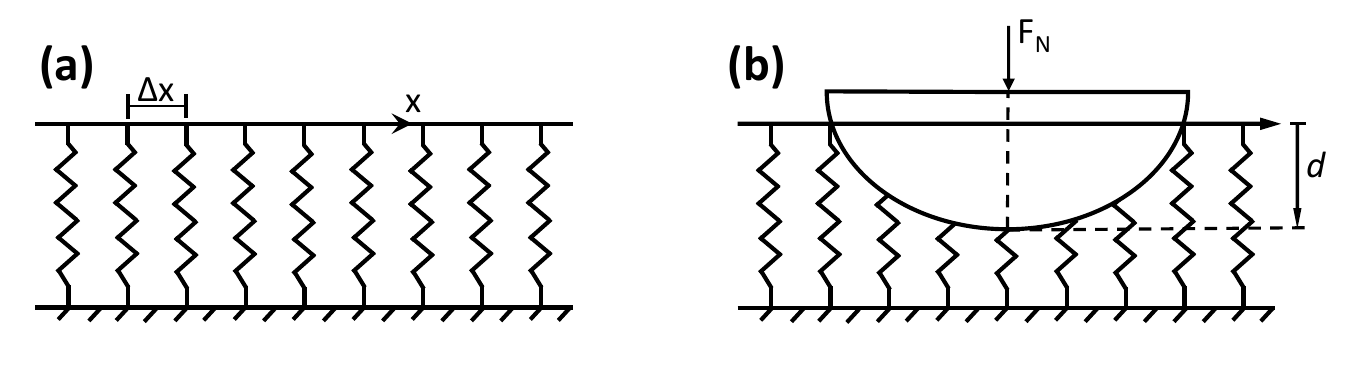}
    \caption{(a)One dimensional elastic foundation. (b) Substitute model for normal contact problem.}
    \label{fig:enter-label}
\end{figure}

\section{Experimental details}
\subsection{Low temperature measurements}
The low temperature measurements were performed in a He-3 cryostat working down to 450 mK. A calibrated ruthenium oxide thermometer and a heater were attached to the sample stage for precise temperature measurement and control. The cerium samples used in these measurements were cut from a high purity (99.9\%) Ce rod from Thermo Fisher Scientific. The sample was placed at the center of a 7 T magnetic field, and all magnetic field measurements were performed with the field applied perpendicular to the sample surface.
\subsection{Point contact spectroscopy measurements}
Point contact spectroscopy measurements were performed using a home-built PCS probe \cite{Shekhar}. The needle-anvil technique was employed to create point contacts between the Cerium sample and metallic needles. Gold(Au), Silver(Ag), and Platinum-iridium(Pt-Ir) needles were used for this purpose. A piezo-controlled nanopositioner was used to precisely control the movement of the needles. Differential conductance ($\frac{dI}{dV}$) vs  $V$  spectra were recorded using lock in amplifier(SR830) based AC modulation technique. In this technique , a dc current  coupled with small ac current is is
passed through point contact. A digital multimeter (Keithley 2000) is used to measure dc output voltage across point contact and AC output is measured by lock in amplifier (SR830).The first harmonic response of lock-in amplifier is proportional to $\frac{dI}{dV}$ which can be used to calculate $\frac{dI}{dV}$ vs $V$ spectrum
\section{Analysis of point Contact spectrosocpy (PCAR) spectra}
We used the Blonder, Tinkham, and Klapwijk (BTK) theory to analyze the PCAR spectra. According to this theory, the current through a point contact between a normal metal and a superconductor (N/S point contact) is given by\par

            I = C \( \int_{-\infty}^{+\infty} [f (E-eV) - f (E)][1+A(E)-B(E)] \,dE \)\par
 
where ,\par
    $f(E)$ is fermi distribution function\par
    $A(E)$ is the probability of Andreev reflection \par
    $B(E)$ is the probability of normal reflection.\par
    PCAR spectra ($\frac{dI}{dV}$ vs $V$) can be simulated using above equation.\par 
    The following formulas were used to calculate $A(E)$ and $B(E)$-
\\ 
For $E< \Delta$:\par
    $ A(E) = \frac{(\Delta/E)^2}{1-\varepsilon(1+2Z^2)^2} $\par
    and $B(E) = 1-A(E)$
\\
For $E>\Delta$:\par
    $ A(E) = \frac{u^2v^2}{\gamma^2} $\par
    and $B(E) = \frac{[u^2-v^2]^2Z^2(1+Z^2)}{\gamma^2}$\par
    where $u^2= 1-v^2=\frac{1}{2}[1+\frac{((E)^2-\Delta^2)^{1/2}}{E}]$, $\gamma^2=([u^2-v^2]Z^2+u^2)^2$, and $\varepsilon= \frac{(E^2-\Delta^2)}{E^2}$
\\
Taking into account the finite lifetime of the quasiparticles at the interface, the BTK theory can be modified by introducing an additional broadening parameter $\Gamma$\cite{Gamma}. The expressions for the coefficients $A(E)$ and $B(E)$ will be modified by replacing $E$ with $E+i\Gamma$. In modified BTK theory, $\Gamma$ along with the parameter Z and the superconducting gap ($\Delta$) are the fitting parameters.\par
\clearpage
\newpage
\section{Origin of large broadening parameter}
Depending on shape of needle , a pressure distribution under the needle is expected which also leads to local distribution of superconducting energy gap.
The broadening parameter $\Gamma$ is commonly attributed to the broadening of the spectrum\cite{Gamma}, however  a distribution of
 gap values can also lead to 
a large $\Gamma$ in BTK analysis \cite{Goutam}.  To demonstrate this, we used a distribution of superconducting energy gaps to simulate a set of spectra, taking mean value of energy gap $<\Delta> =$ 2 meV and setting $\Gamma=0$. The normalized conductance $(\frac{dI}{dV})_N$ vs $V$ curves are computed from $I= \sum_iI_{BTK}(T,\Delta_i,\Gamma=0,V)$, where $I_{BTK}$ is BTK equation for current and sum runs over distribution of gap values. Normalised conductance $(\frac{dI}{dV})_N$ vs $V$ curves for varying distribution of gap values are shown in Figure S2, colored circles represents the simulated conductance spectra assuming different distribution of $\Delta$ at T = $2K$ with $\Gamma$ = 0. The solid black lines represents the best fits of the simulated curves using single BTK function taking  $\Gamma$ as additional fitting parameter. We can see that broadening parameter $\Gamma$ increases with distribution of energy of gap, showing that $\Gamma$ is also measure of variation in gap values.

\begin{figure}[h!]
    \centering
    \includegraphics[width=0.7\linewidth]{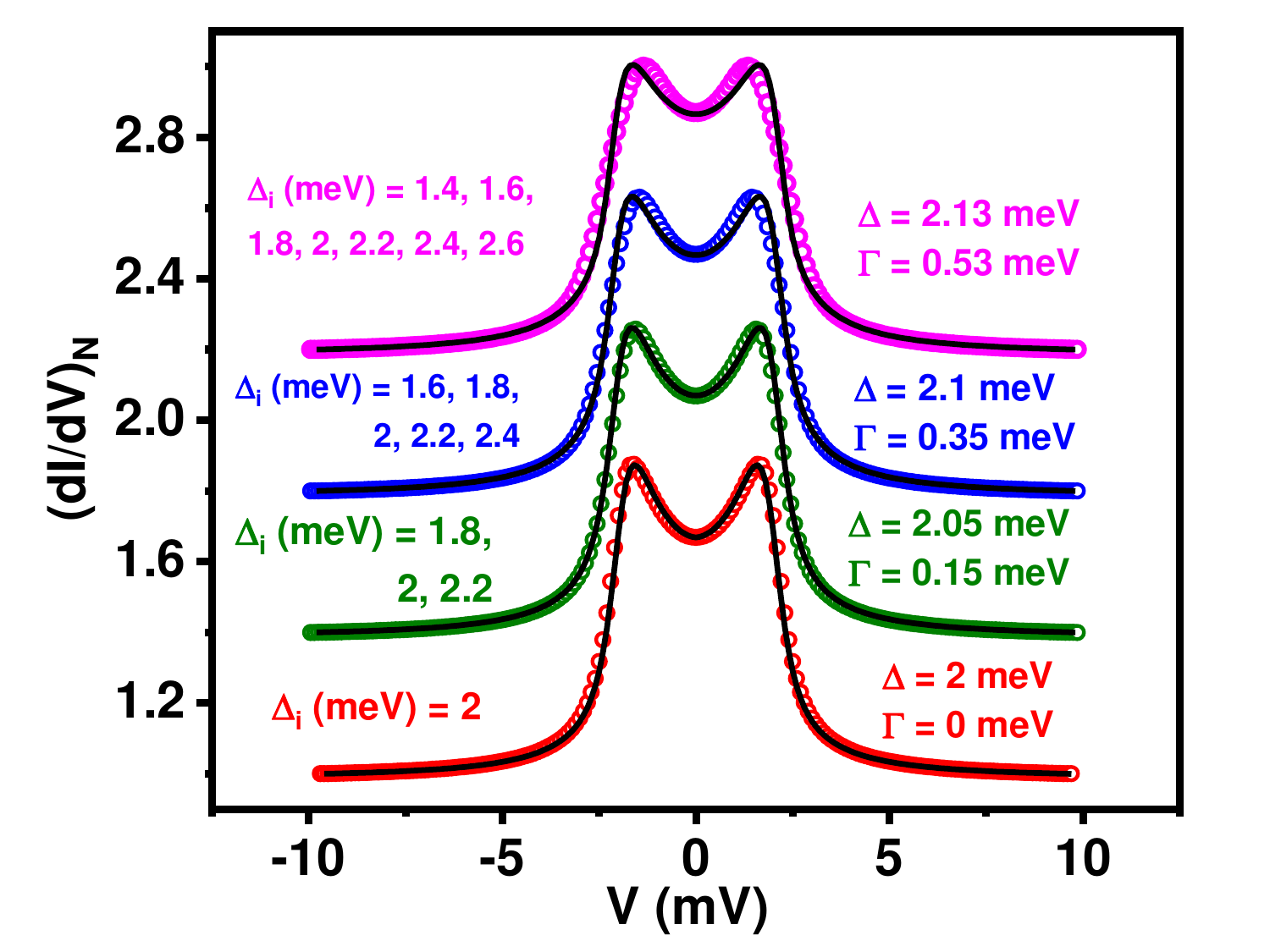}
    \caption{Colored circles represent the simulated conductance spectra assuming different distribution of $\Delta$ at 2K with broadening parameter ($\Gamma$) = 0 meV. The solid black line over the colored circles represents the best-fitting curve with $\Gamma$ included.}
    \label{high_gamma}
\end{figure}
\clearpage

\section{Size of point contacts}
\justifying
Size of different point contacts can be estimated by using normal state resistance following Wexler's formula given by:\par
\centering
$R_{PC} = \frac{2h/e^2}{(ak_F)^2} + \Gamma(l/a)\frac{\rho(T)}{2a}$\par
\justifying
where $\Gamma(l/a)$ is the numerical factor close to unity , $a$ is the contact diameter, $l$ is mean free path of electrons, $2h/e^2$ is quantum resistance equal to 50 $k\Omega$, $\rho(T)$ is the resistivity of material at temperature T, $k_F$ is Fermi wave vector. The formula for contact resistance consists of two terms: the first term represents the Sharvin's resistance, and the second term corresponds to the Maxwell's resistance. In the ballistic regime, where the contact diameter is much smaller than the elastic mean free path, the resistance is primarily determined by the Sharvin's resistance. However, in the thermal regime, where contact diameter is larger then inelastic mean free path, the Maxwell's resistance dominates. In intermediate regime, where point contact diameter lies between ballistic and thermal regime , both Sharvin's and Maxwell's resistance contribute.
\begin{figure}[h!]
 \centering
 \includegraphics[scale = 0.8]{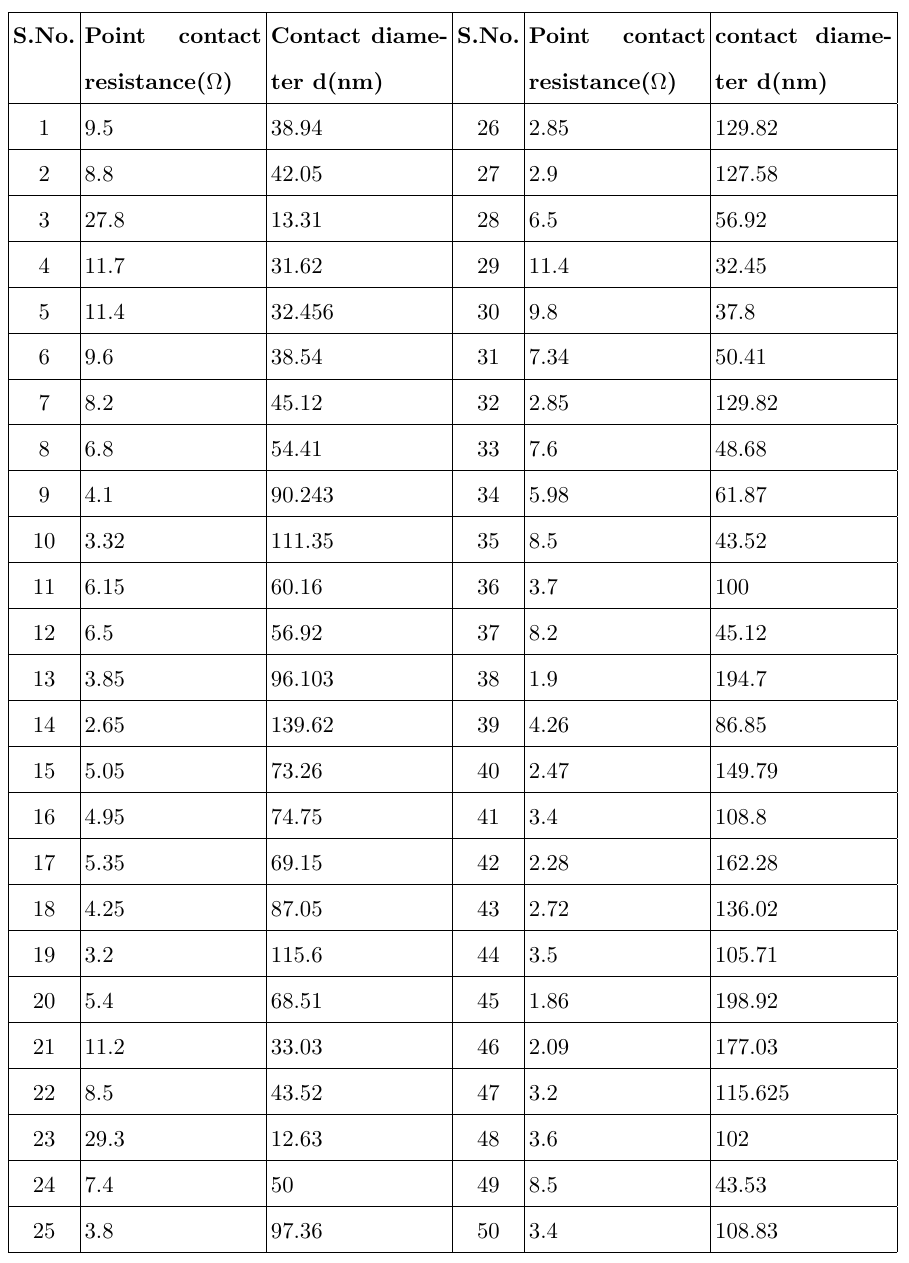}
 \caption{A list of normal state resistance and estimated diameter of different point contacts.}
\end{figure}
\clearpage
\newpage
\begin{figure}[h!]
 \centering
 \includegraphics[scale = 0.8]{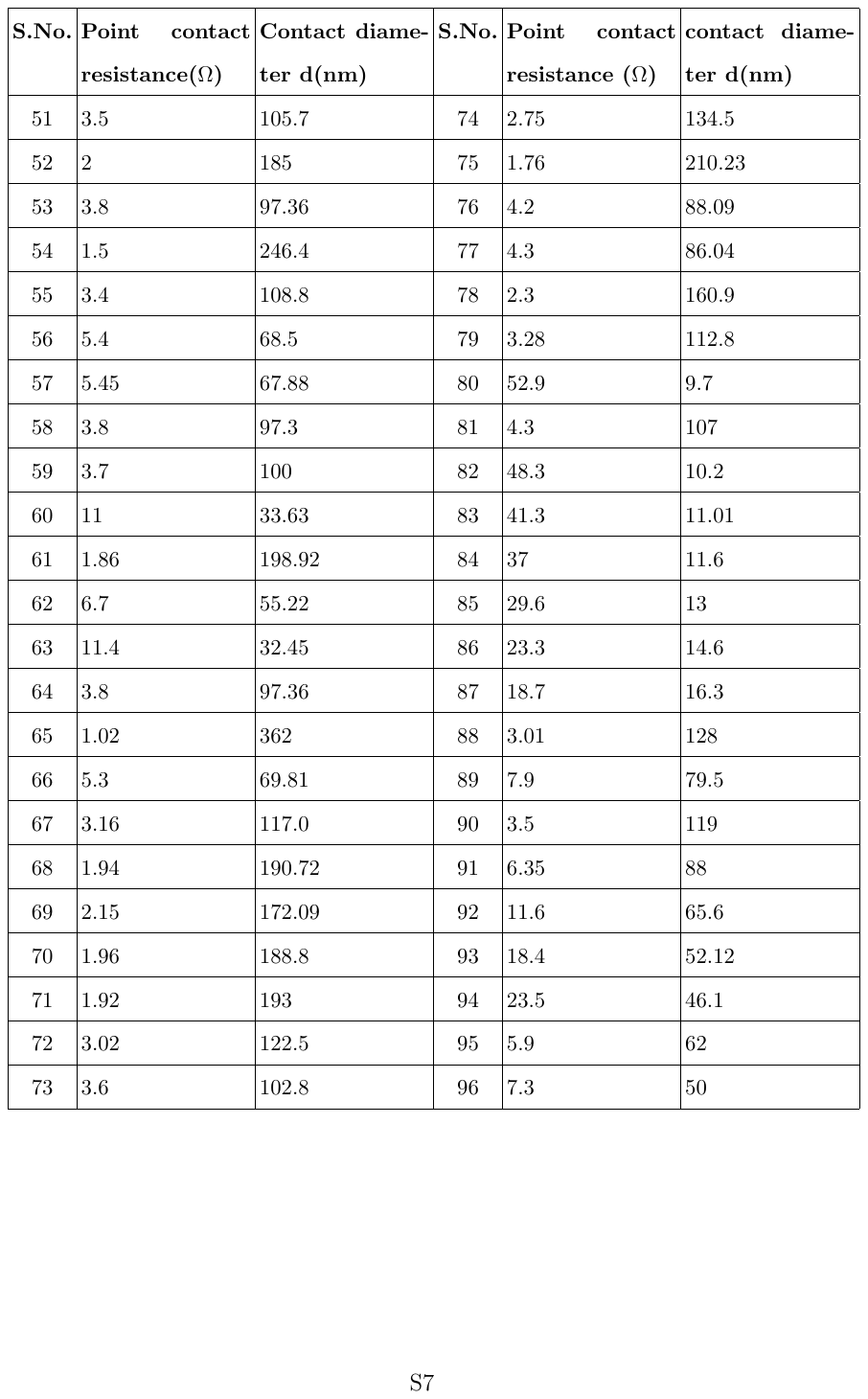}
 \caption{A list of normal state resistance and estimated diameter of different point contacts.}
\end{figure}
\clearpage
\newpage
\justifying
\section{Raw data for R vs T at different magnetic fields}
\begin{figure}[h!]
    \centering
    \includegraphics[width=0.7\linewidth]{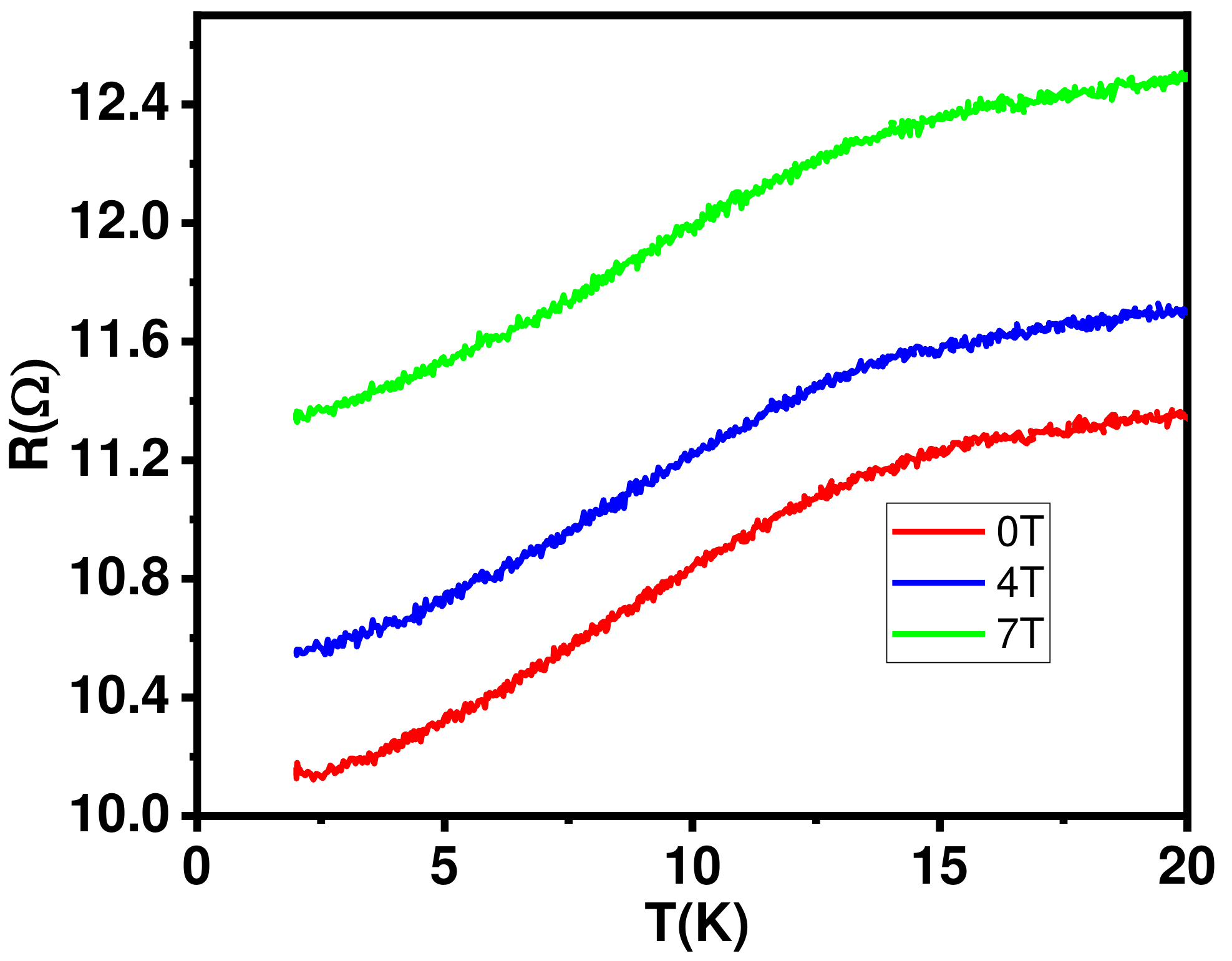}
    \caption{The raw data corresponds to Figure 1(e) in main manuscript.}
    \end{figure}
\justifying
\section{Determination of critical temperature T$_C$}
\justifying
A broad transition into superconducting state can be seen in the R-T data for different point contacts. We have plotted the slopes of R-T curves both above and below the onset of transitions. The temperature corresponding to the point of intersection of two slopes is taken as critical temperature T$_c$ for corresponding point contact. We performed a temperature dependence study of the spectra in different regimes until the spectral features related to superconductivity completely disappeared. The temperature at which the spectral features completely disappeared is taken as T$_c$.
\clearpage
\newpage

\section{Additional data}
\subsection{Temperature dependence of conductance spectra using Pt-Ir needle}
\begin{figure}[h!]
 \centering
 \includegraphics[width = 1\linewidth]{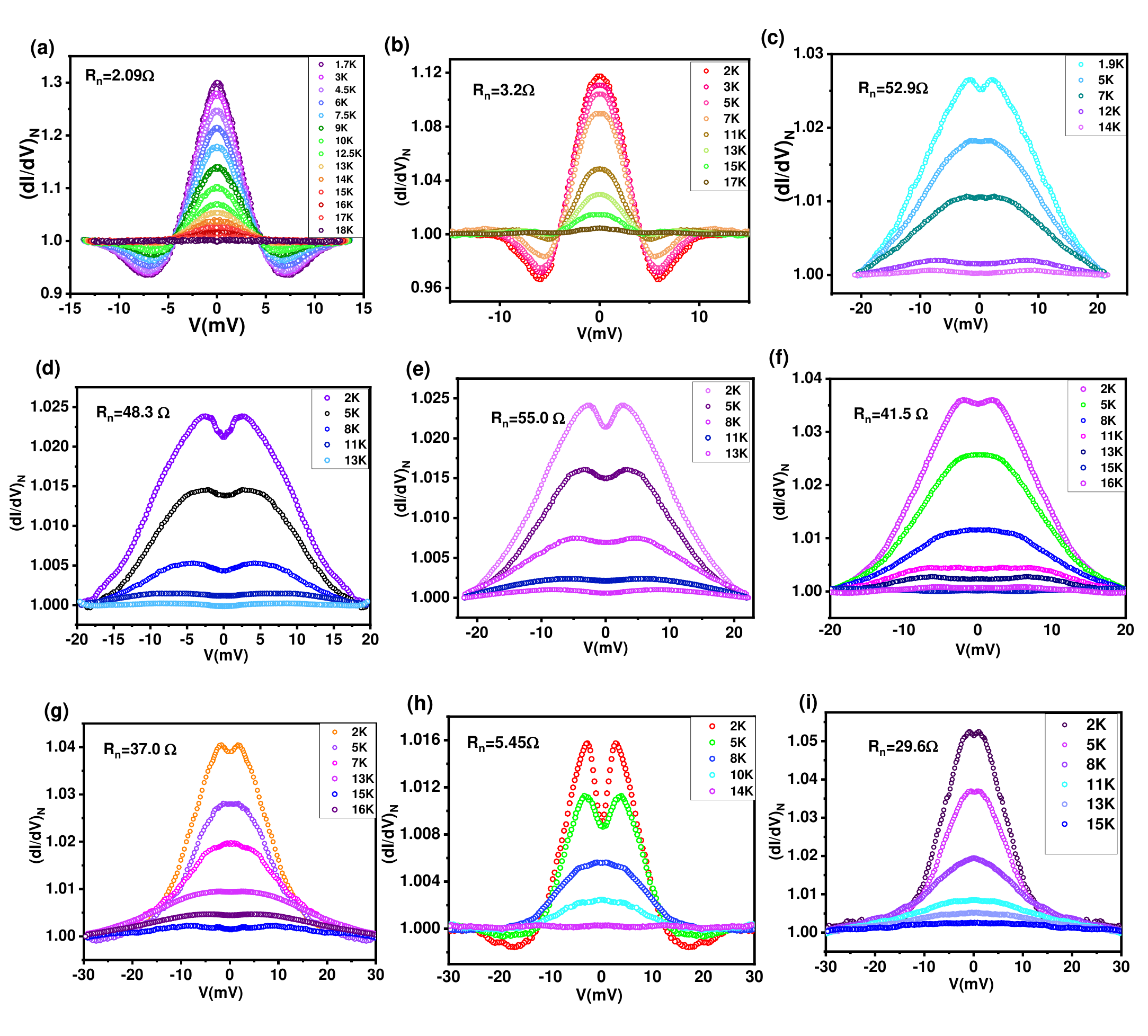}
 \caption{Temperature dependence of the conductance spectra at different point contacts using Pt-Ir needle.}
\end{figure}
\begin{figure}[h!]
    \centering
    \includegraphics[width = 1\linewidth]{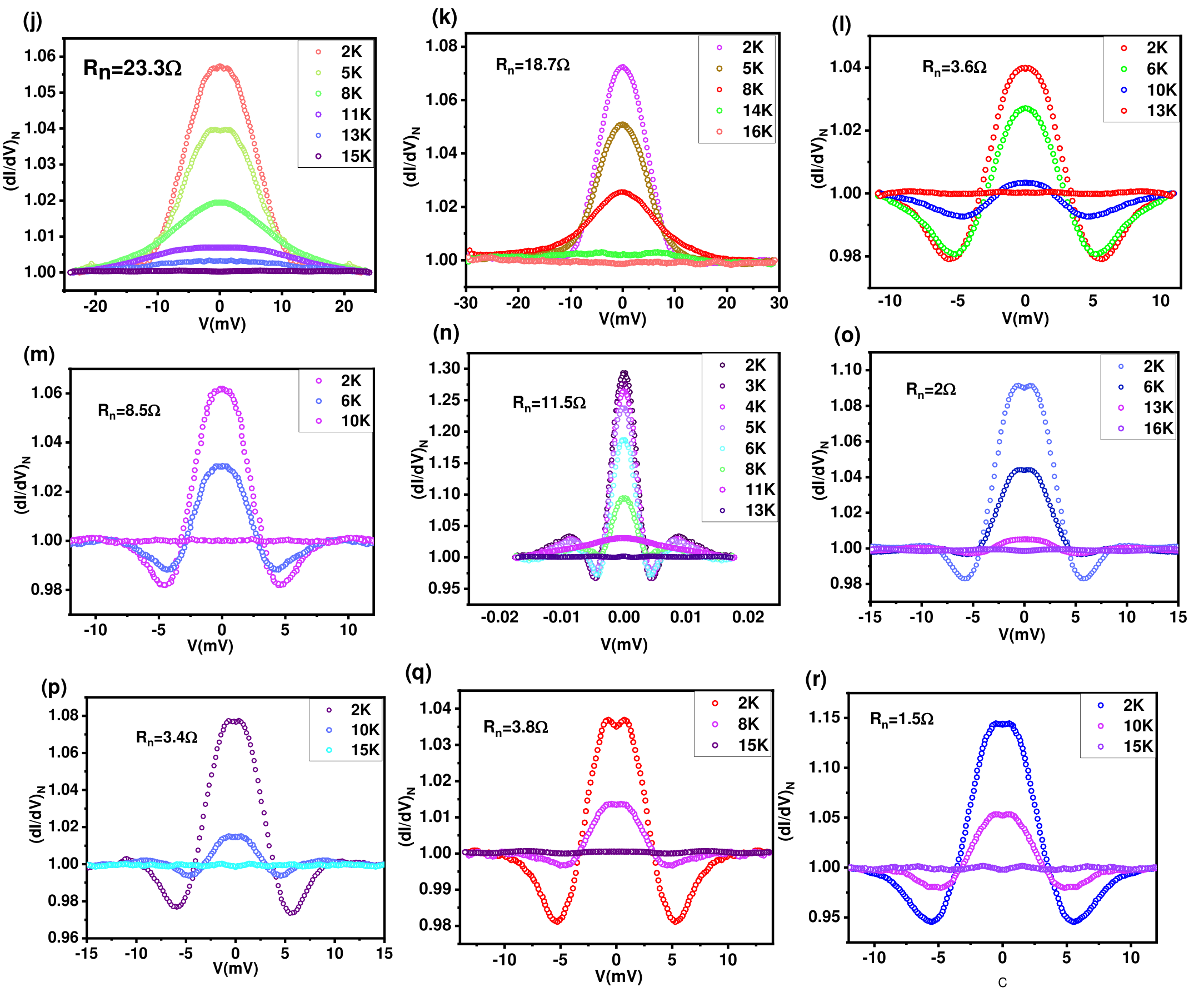}
 \caption{Temperature dependence of the conductance spectra at differentt point contacts using Pt-Ir needle}
\end{figure}
\clearpage
\newpage
\subsection{Temperature dependence of conductance spectra using Silver needle }
\begin{figure}[h!]
    \centering
    \includegraphics[width = 1\linewidth]{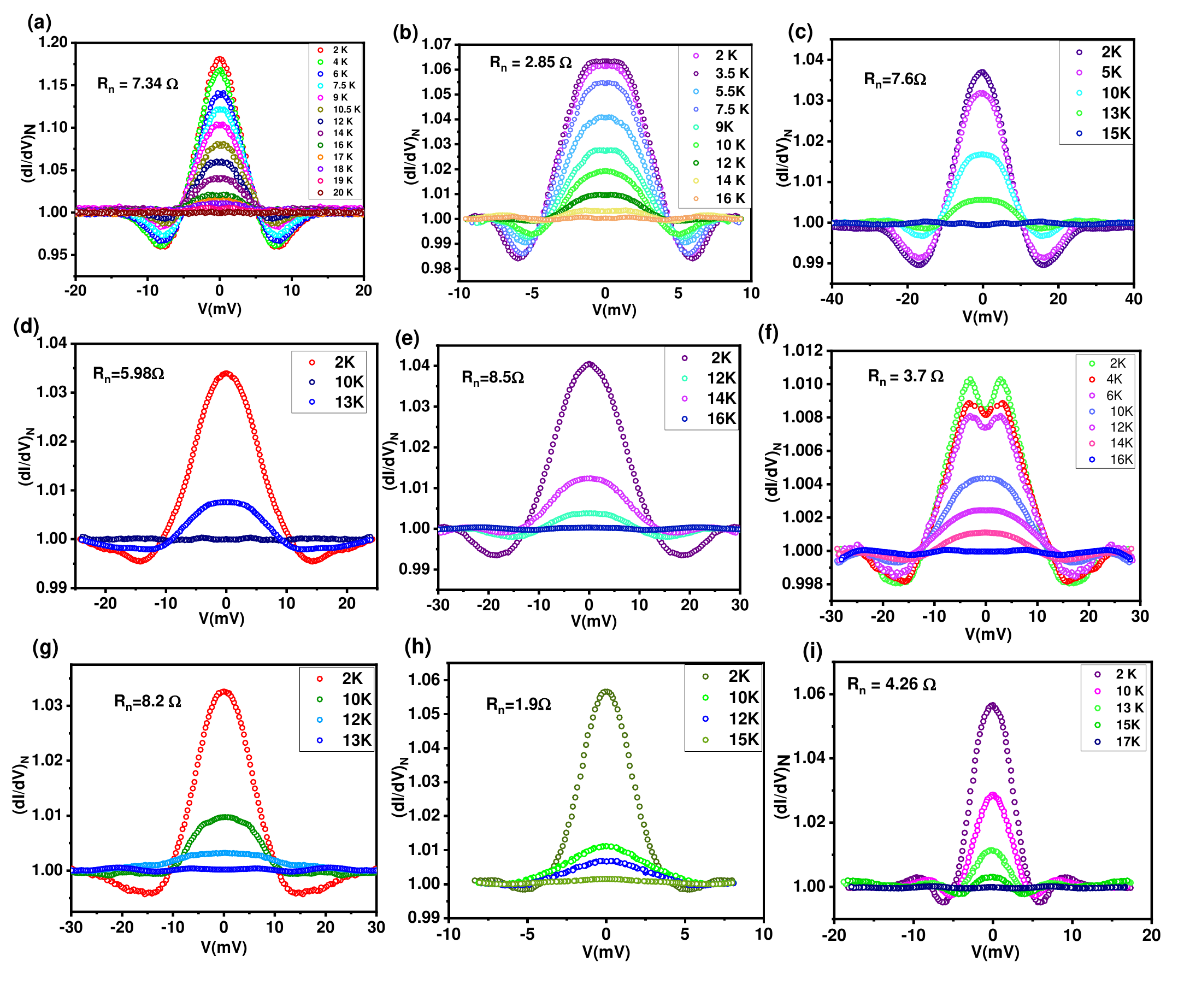}
    \caption{Temperature dependence of the conductance spectra at different point contacts using silver needle.}
    
\end{figure}
\begin{figure}[h!]
    \centering
    \includegraphics[width = 1\linewidth]{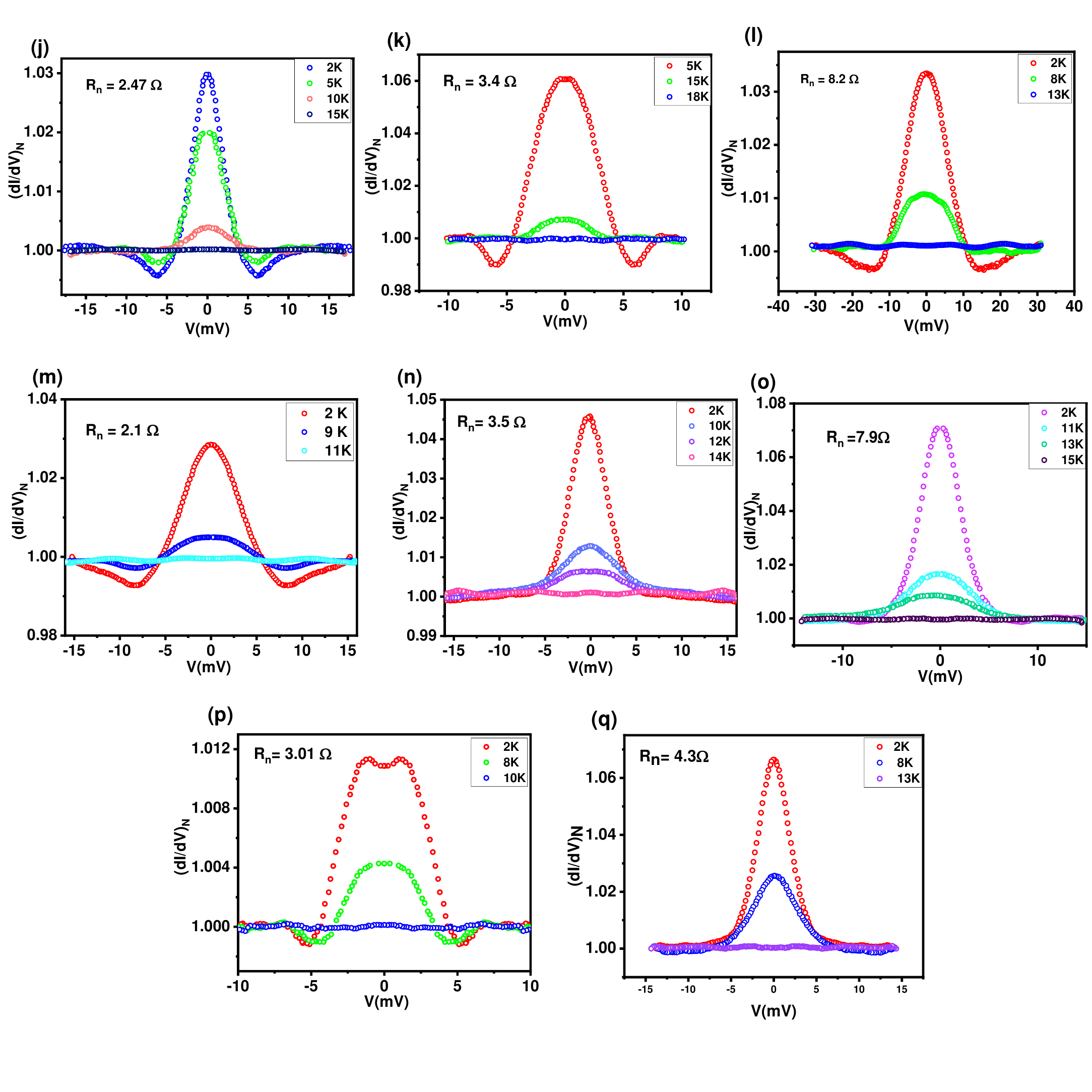}
    \caption{Temperature dependence of the conductance spectra at different point contacts using silver needle.}
    
\end{figure}
\clearpage
\newpage
\subsection{\textbf{Resistance(R) vs temperature(T) of point contacts between Cerium and gold needle }}
\begin{figure}[h!]
    \centering
    \includegraphics[scale=0.47]{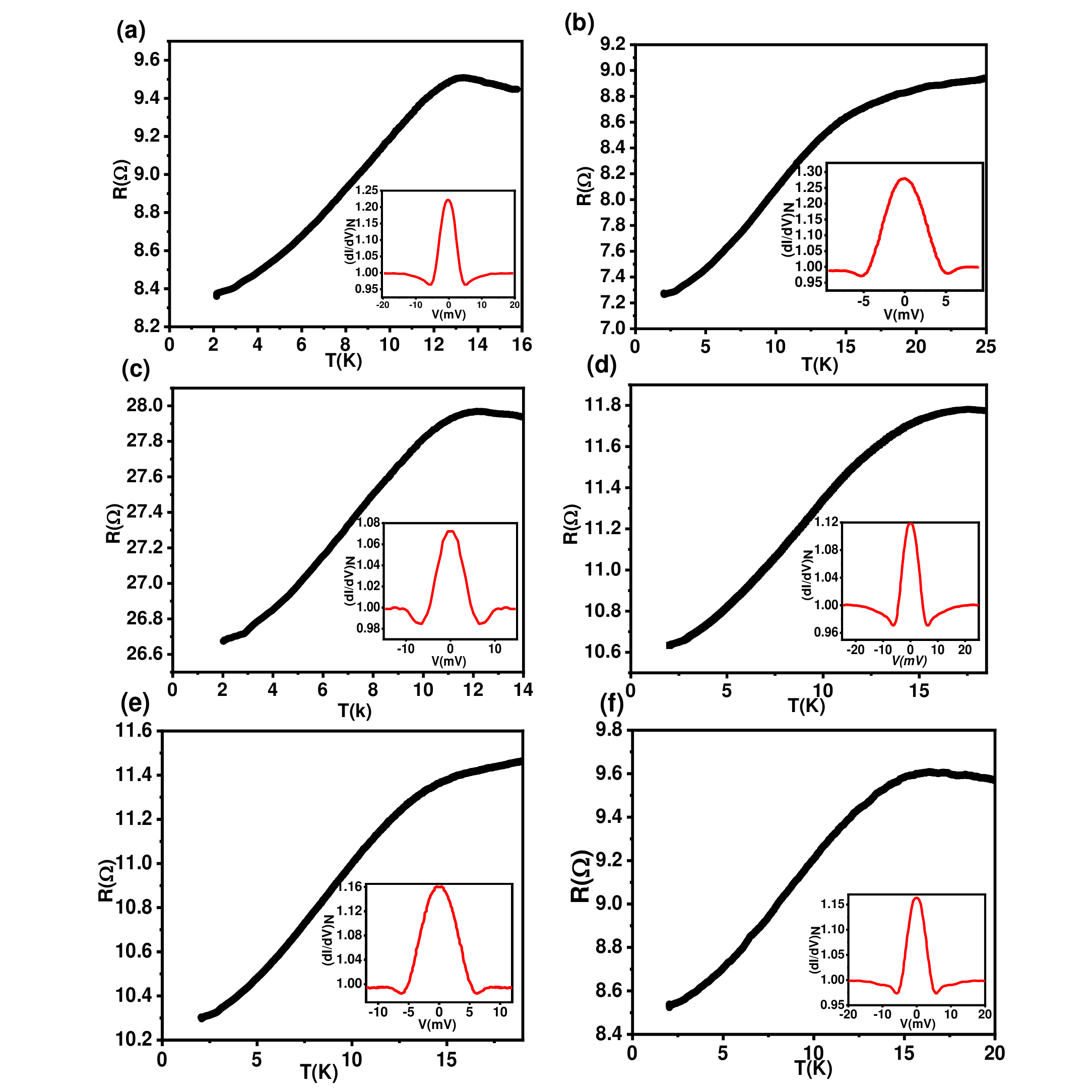}
    \caption{Resistance (R) vs. temperature (T) of different point contacts between Ce and Au needle. The inset displays the corresponding conductance spectra in thermal regime at 2 K.}
    \end{figure}
\begin{figure}[h!]
    \centering
    \includegraphics[scale=0.47]{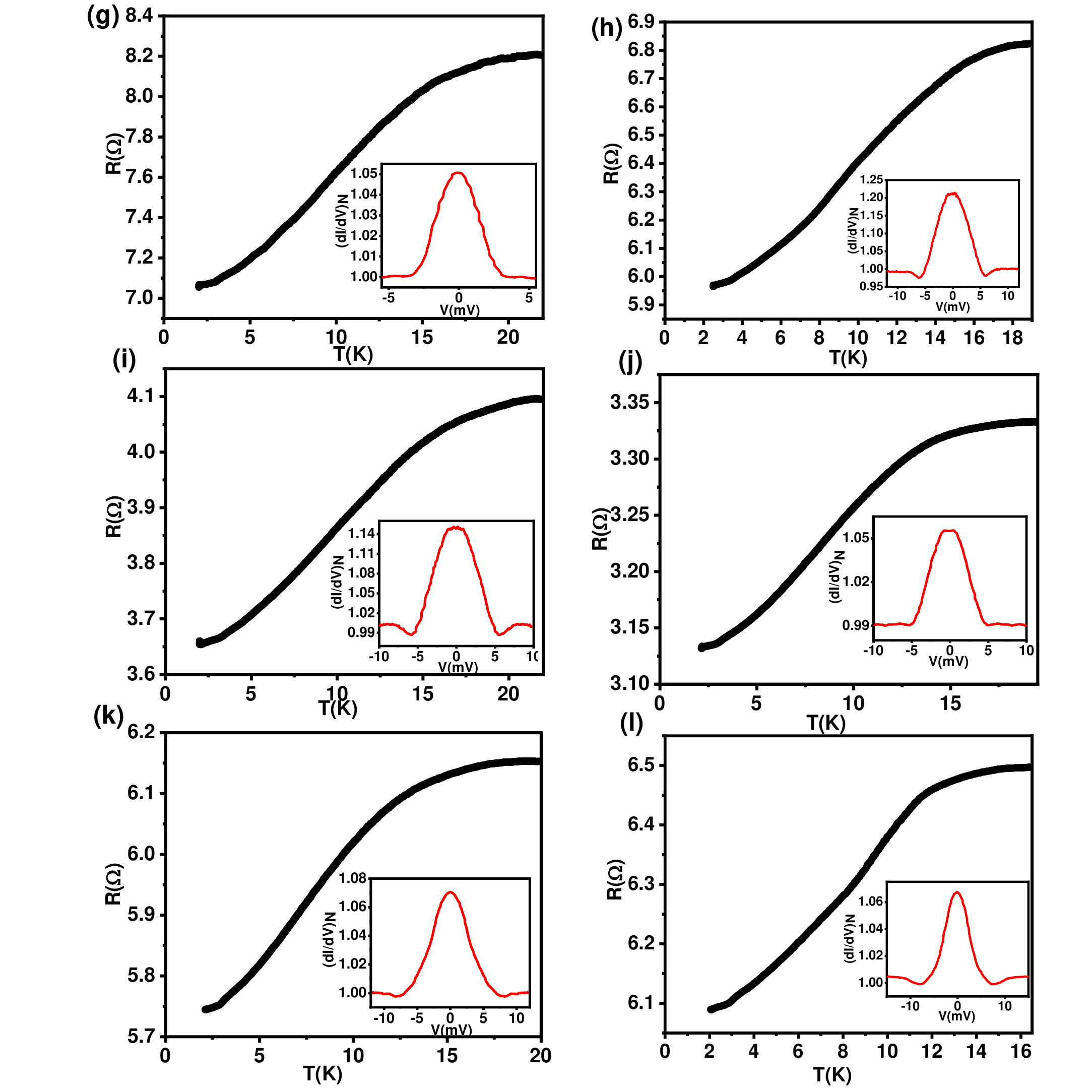}
    \caption{Resistance (R) vs. temperature (T) of different point contacts between Ce and Gold (Au) needle. The inset displays the corresponding conductance spectra in thermal regime at 2 K.}
    \end{figure}
\begin{figure}[h!]
    \centering
    \includegraphics[scale=0.47]{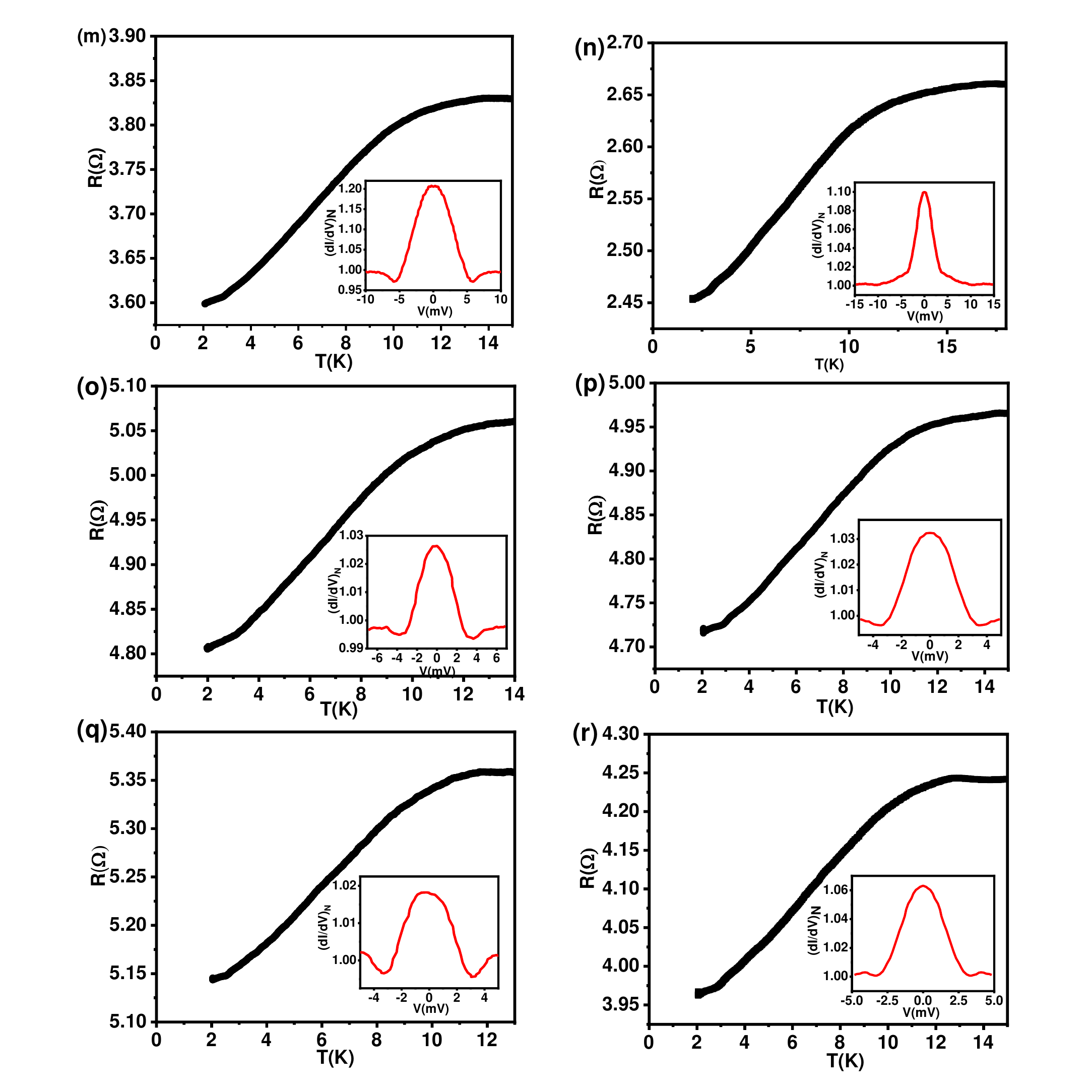}
    \caption{Resistance (R) vs. temperature (T) of different point contacts between Ce and Gold (Au) needle. The inset displays the corresponding conductance spectra in thermal regime at 2 K.} 
\end{figure}
\begin{figure}[h!]
    \centering
    \includegraphics[scale=0.47]{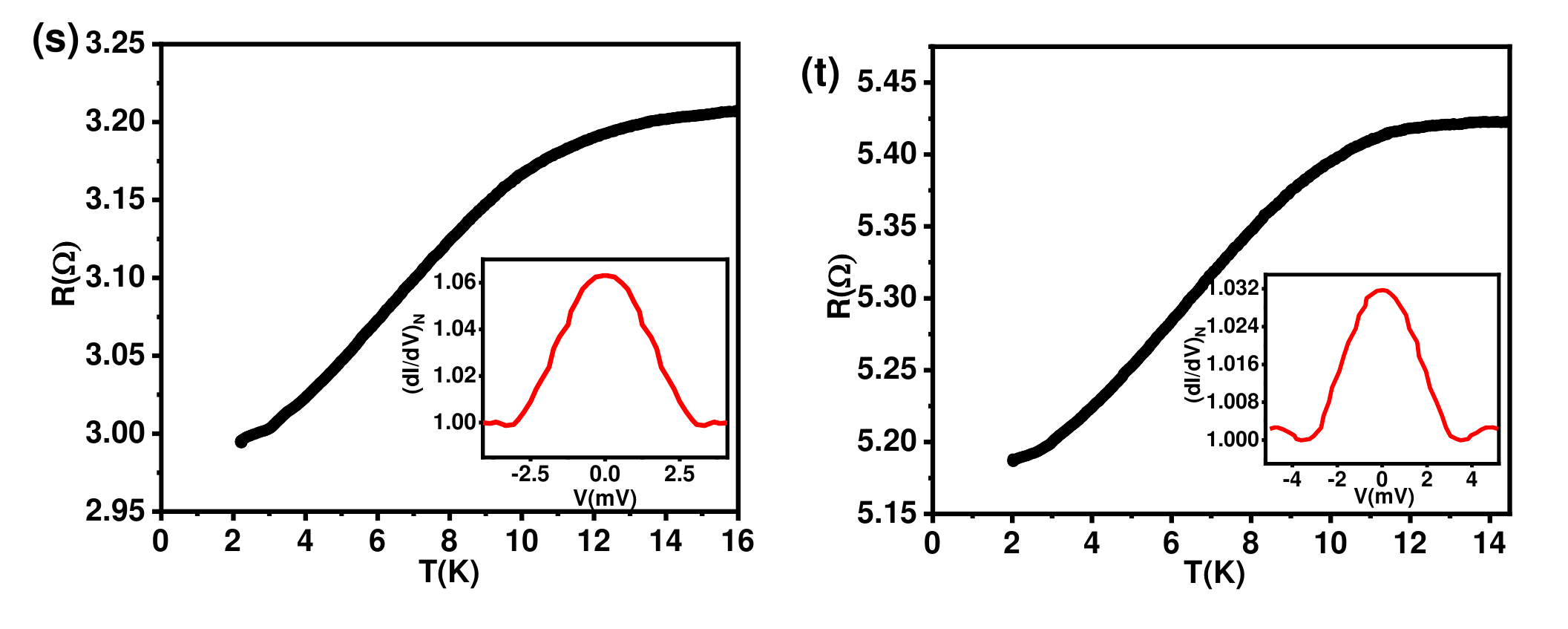}
    \caption{Resistance (R) vs. temperature (T) of different point contacts between Ce and Gold (Au) needle. The inset displays the corresponding conductance spectra in thermal regime at 2 K.} 
\end{figure}
\clearpage
\subsection{Magnetic field dependence of Resistance (R) vs Temperature (T) }
\begin{figure}[h!]
    \centering
    \includegraphics[scale=0.47]{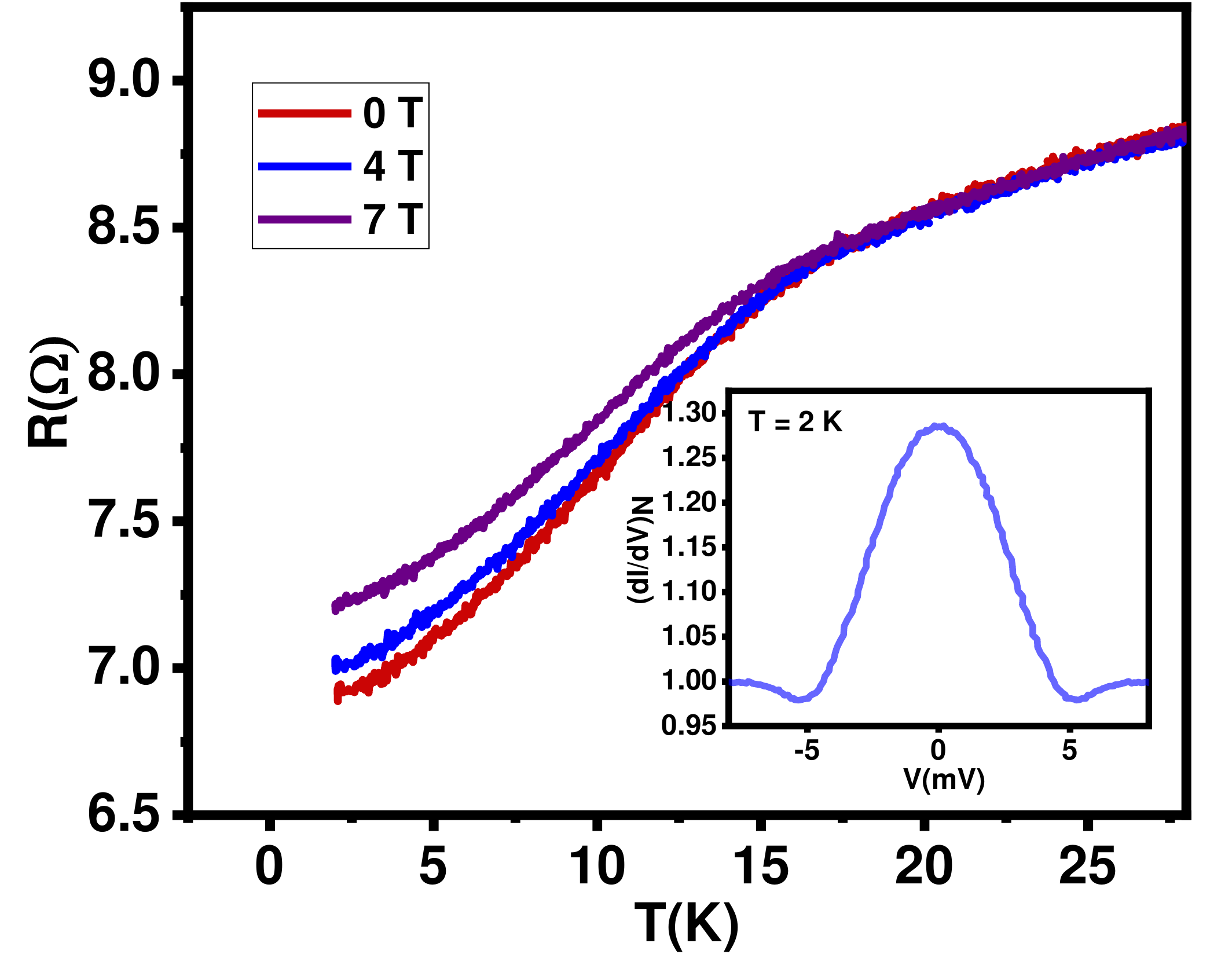}
    \caption{R vs. T of the point contact at different magnetic fields (H) up to 7 T. The inset shows the corresponding conductance spectrum.}
\end{figure}
\clearpage
\newpage
\section{Point contact measurements on Scandium }
\begin{figure}[h!]
    \centering
    \includegraphics[width = 1\linewidth]{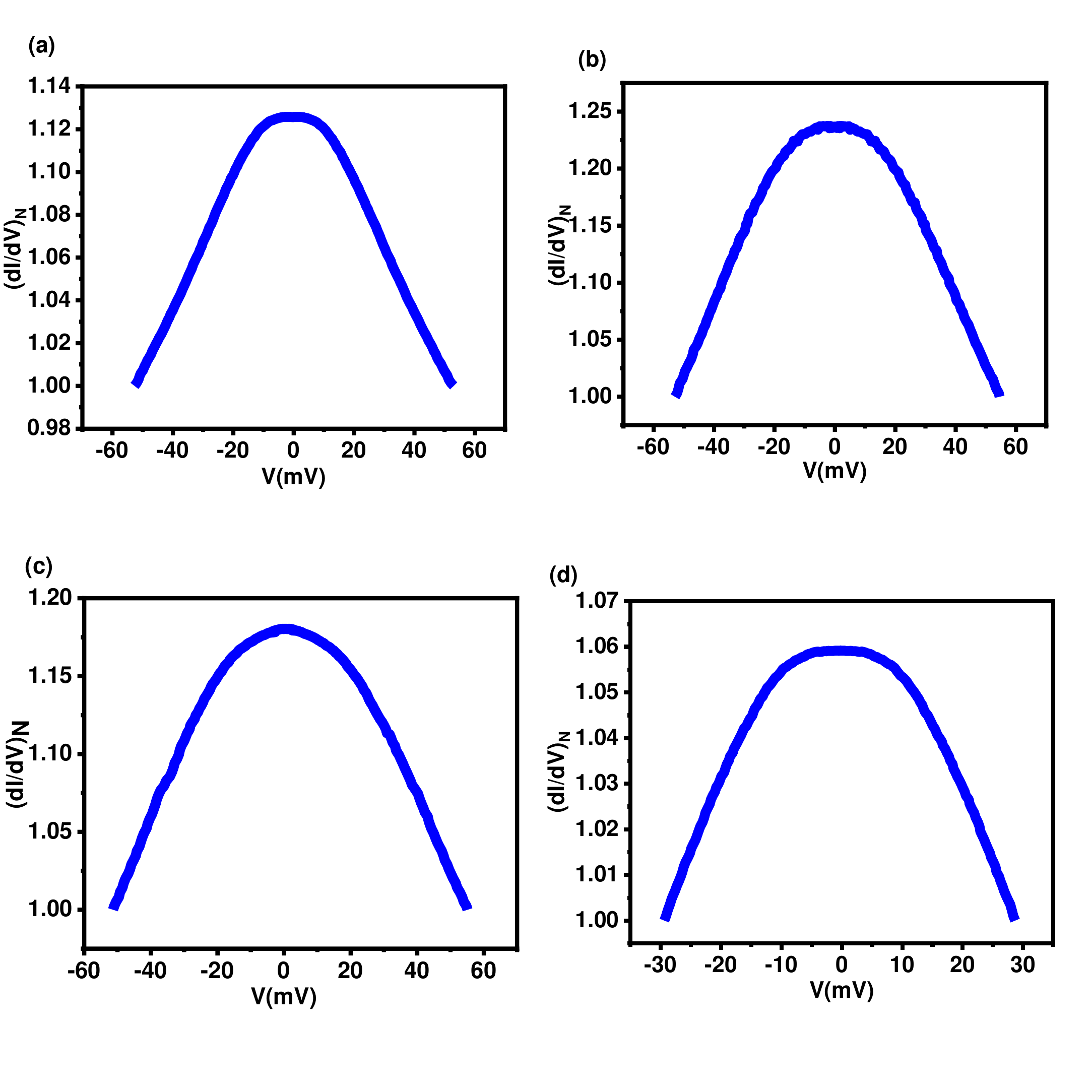}
    \caption{Conductance spectra for different point contacts between Scandium and Pt-Ir needle at 2 K.}
    \label{fig:enter-label}
\end{figure}
    \clearpage
    \newpage
\section{Point contact measurements on Lanthanum }
\begin{figure}[h!]
    \centering
    \includegraphics[width = 1\linewidth]{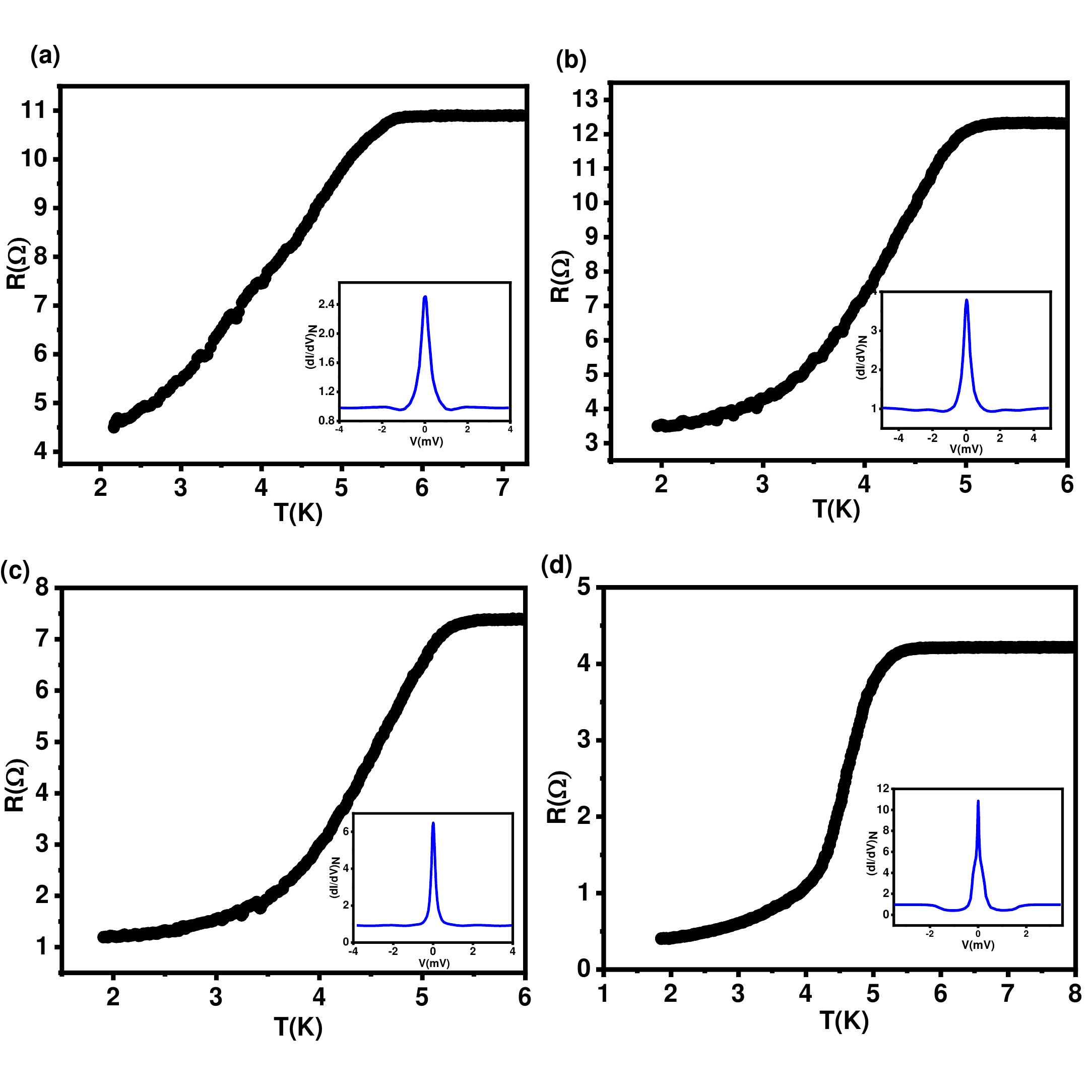}
    \caption{Resistance (R) vs. temperature (T) of different point contacts between Lanthanum and Pt-Ir needle. The inset displays the corresponding conductance spectra in thermal regime at 2 K.}
    \end{figure}
   \clearpage
   \newpage